\documentclass[3p,times]{elsarticle}

\usepackage{ecrc}

\usepackage{epstopdf}

\volume{00}

\firstpage{1}

\journalname{Nuclear Physics A}

\runauth{Rupa Chatterjee et al.}

\jid{nupha}

\jnltitlelogo{Nuclear Physics A}

\usepackage{graphicx}
\usepackage{amsmath,amssymb}

\begin{document}

\begin{frontmatter}



\title{Thermal photon $v_3$ at LHC from fluctuating initial conditions}

\author[label1]{Rupa Chatterjee}
\ead{rupa@vecc.gov.in}

\author[label1]{Dinesh K. Srivastava}
\author[label2,lebel3]{Thorsten Renk}

\address[label1]{Variable Energy Cyclotron Centre, 1/AF, Bidhan Nagar, Kolkata-700064, India}
\address[label2]{Department of Physics, P.O.Box 35, FI-40014 University of Jyv\"askyl\"a, Finland}
\address[lebel3]{Helsinki Institute of Physics, P.O.Box 64, FI-00014 University of Helsinki, Finland}
\begin{abstract}
We calculate the triangular flow parameter $v_3$ of thermal photons for 0--40\% central collisions of Pb nuclei at LHC using an event-by-event hydrodynamic model with fluctuating initial conditions. Thermal photon $v_3$  with respect to the the participant plane angle is found to be positive and significant compared to the elliptic flow parameter $v_2$ of thermal photons. In addition, photon  $v_3$ as a function of $p_T$ shows similar qualitative nature to photon $v_2$ in the region $1< p_T <6$ GeV/$c$.  We argue that while $v_3$ originates from $\epsilon_3$ deformations of the initial state density distribution, fast buildup of radial flow due to fluctuations is the main driving mechanism for the observed large value.  
\end{abstract}

\begin{keyword}
Triangular flow, event-by-event hydrodynamics, thermal photons

\end{keyword}

\end{frontmatter}



\section{Introduction}
\label{intro}
The observation of a significant triangular flow parameter $v_3$ of charged hadrons is considered one of the most interesting results from recent heavy ion experiments at the Relativistic Heavy Ion Collider (RHIC) and at the Large Hadron Collider (LHC) energies~\cite{phenix_flow,alice_flow,flow_atlas}. This large value  is attributed to the fluctuations in the initial  QCD matter density distribution~\cite{alver,nex,flow_lhc}. As thermal emission of photons is known to be sensitive to the initial temperature of the system, the study of photon $v_3$ can provide valuable information about initial state fluctuation and and its evolution complementary to hadronic observables.

Hydrodynamic model calculation with event-by-event (e-by-e) fluctuating initial conditions (IC) explain the data for hadronic spectra and elliptic flow well upto a large $p_T$ even for most central collisions~\cite{hannu}. It has also been shown that fluctuations  in the IC enhance the production of thermal photons significantly for $p_T >$ 2 GeV/$c$ compared to a smooth initial-state-averaged profile and a better approximation of the direct photon data is obtained in that $p_T$ region~\cite{chre1,chre2}.
In a recent study we have demonstrated that e-by-e fluctuating IC result in much larger elliptic flow of thermal photons  than a smooth IC for $p_T>$ 2 GeV/$c$ both at RHIC and at the LHC energies~\cite{chre3}. However, we see that the effect of IC fluctuations in not sufficient to explain the direct photon $v_2$ data and our elliptic flow  results still underestimate the data by a large margin~\cite{chre3,alice_v2}.
Since a nonzero $v_3$ arises only if there are fluctuations in the initial matter density distribution, photon $v_3$ can provide information about the fluctuating initial state more directly than the elliptic flow parameter~\cite{csr,uli_v3}. 

\section{Triangular flow of thermal photons from e-by-e hydrodynamics}
We calculate thermal photon $v_3$ for 0--40\% central collisions of Pb nuclei at LHC using the (2+1) dimensional e-by-e hydrodynamic framework developed in~\cite{hannu}.
This ideal hydrodynamic model with fluctuating IC is in good agreement with charged particle spectra and elliptic flow~\cite{hannu} as well as spectra and elliptic flow of thermal photons at RHIC and at the LHC energies~\cite{chre1,chre2,chre3}. We use state of the art photon rates (complete leading order plasma rates from~\cite{amy} and hadronic photons from~\cite{trg}) integrated over the hydrodynamical evolution to calculate the spectra and the anisotropic flow parameters. The initial formation time $\tau_0$ of the plasma is assumed to be 0.14 fm/c and the fluctuation size parameter $\sigma$ as 0.4 fm (see~\cite{csr} and references therein for details). 

\begin{figure}
\begin{center}
\includegraphics*[width=13cm,clip=true]{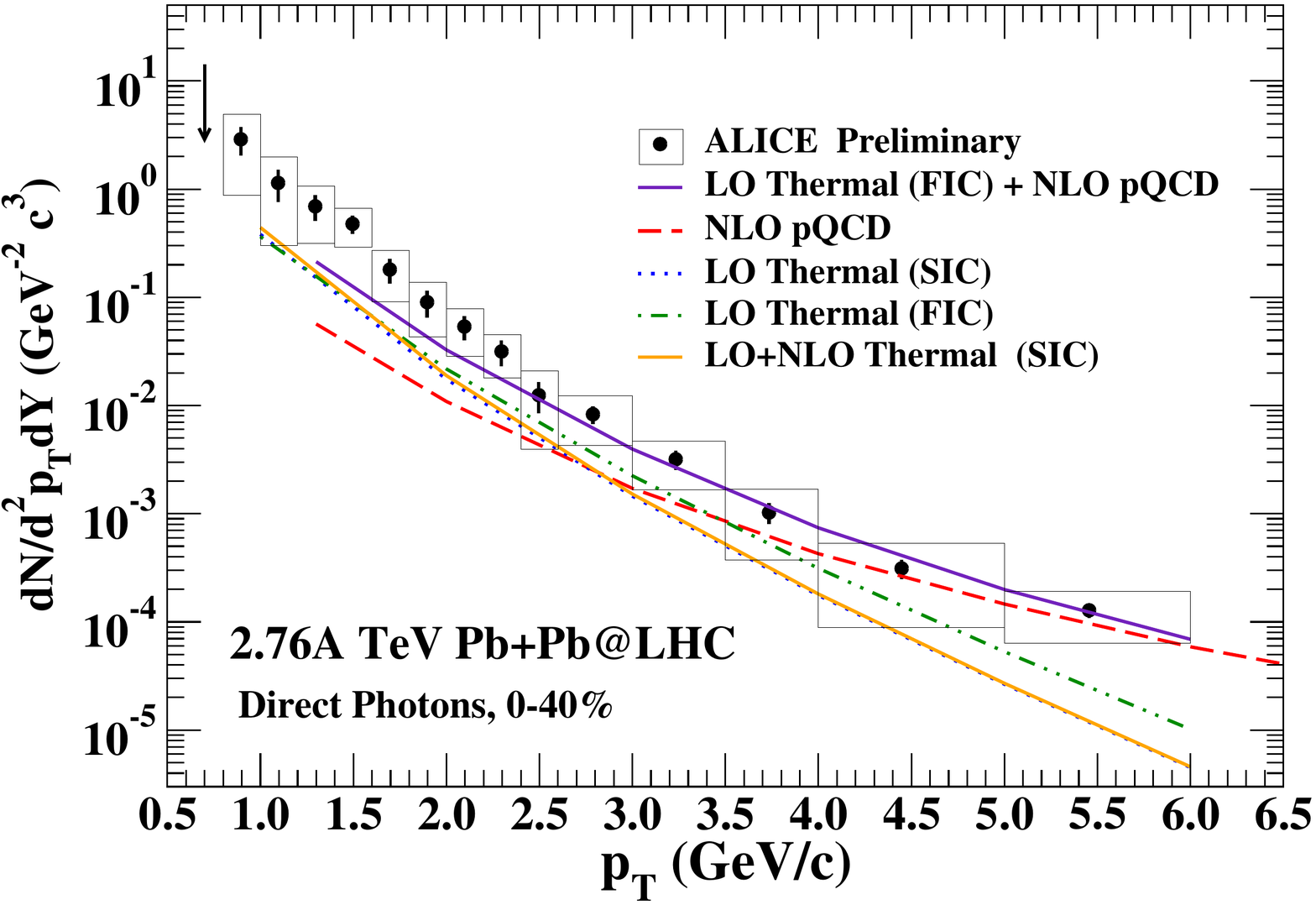}
\caption{Thermal photon $p_T$ spectra  from smooth (SIC) and fluctuating (FIC) initial conditions along with ALICE direct photon data~\cite{alice} for 0--40\% central collision of Pb nuclei at LHC. $p_T$ spectra calculated using leading order (LO) plasma rates~\cite{amy}, next-to-leading order (NLO) plasma rates~\cite{jaccopo} and NLO pQCD photons are
also plotted for comparison [see ~\cite{csr}].}
\label{spec_lhc}
\end{center}
\end{figure}
Figure~\ref{spec_lhc} shows thermal photon spectra from e-by-e hydrodynamics along with ALICE direct photon data~\cite{alice} for 0--40\% central collisions of Pb nuclei at LHC~\cite{csr}. Similar to RHIC, the $p_T$ spectrum from fluctuating initial conditions (FIC) is found to be much  harder than the result obtained from smooth initial conditions (SIC) in the region $p_T >$ 2 GeV/$c$.
Thermal photons from fluctuating IC along with next-to-leading order (NLO) pQCD photons (which starts dominating the direct photon spectrum for $p_T > $ 3.5 GeV/$c$) explain the ALICE data well for $p_T> 2$  GeV/$c$. However, for $p_T <$ 2 GeV/$c$ our result  underestimates the data which can only be explained if the contribution from the hadronic phase is increased by an order of magnitude. 

We see that the inclusion of the NLO  plasma rates~\cite{jaccopo} to our calculation enhance the production of thermal photons by a factor of  about 10--20\% in the region $p_T <$ 2 GeV/$c$. For $p_T>$ 2 GeV/$c$ however, the difference between the complete LO  and NLO results is not significant as shown in Figure~\ref{spec_lhc}. In addition, e-by-e calculation of anisotropic flow parameter using the NLO plasma rate is numerically expensive process.  Hence, in practice we calculate thermal photon $v_3$ using the complete LO plasma rates~\cite{amy} which is an excellent approximation.

\begin{figure}
\begin{center}
\includegraphics*[width=13.cm,clip=true]{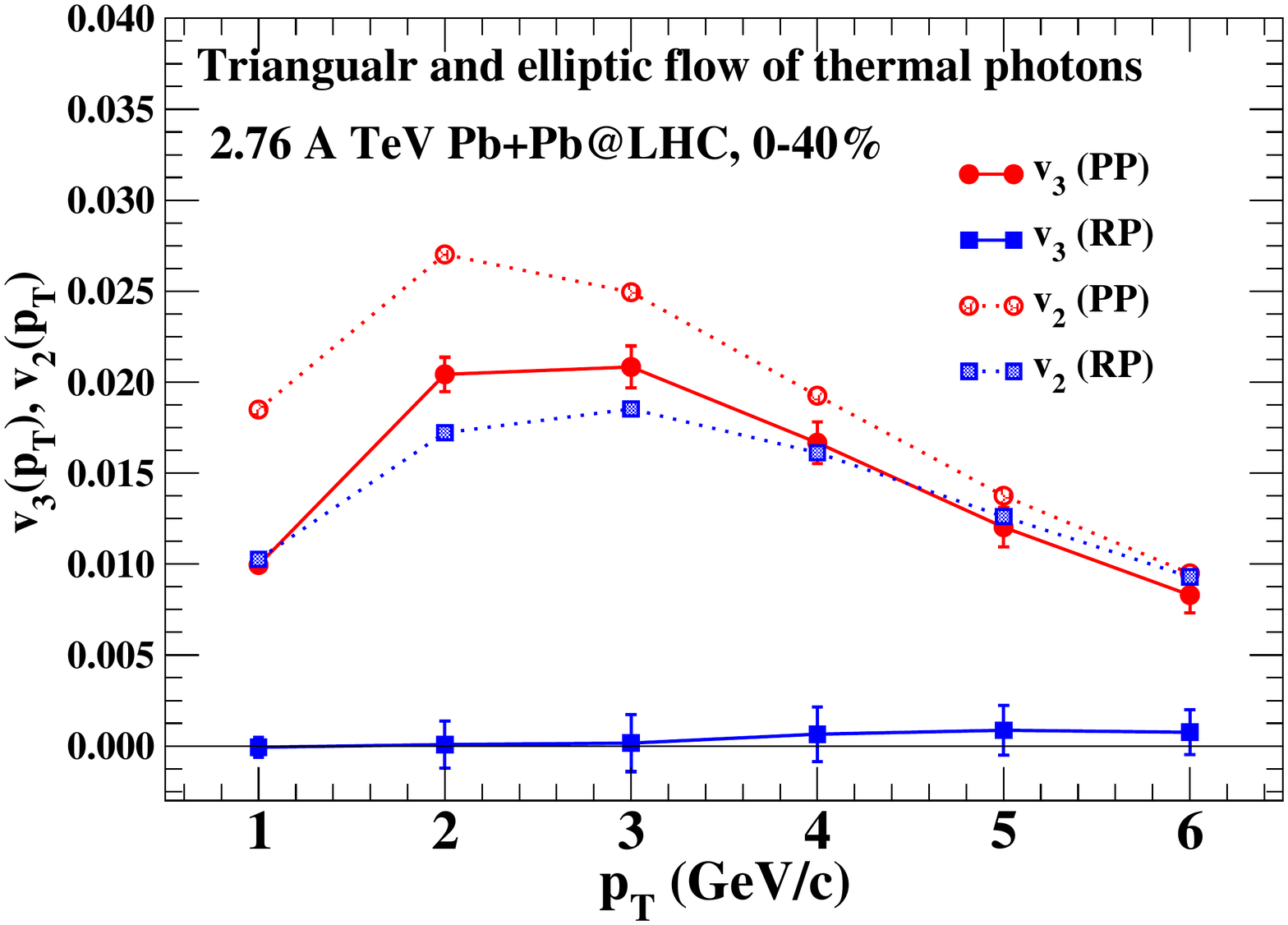}
\caption{Thermal photon $v_3$ with respect to participant plane (PP) and reaction plane (RP) for 2.76A TeV Pb+Pb collisions at LHC and for 0--40\% centrality bin along with photon $v_2$ results~\cite{csr}. }
\label{v3}
\end{center}
\end{figure}

The triangular flow parameter $v_3$ of thermal photons for 0--40\% central collisions of lead nuclei at 2.76A TeV at LHC is shown in figure~\ref{v3}. $v_3$ is calculated with respect to the participant plane (PP) by taking average over a sufficiently large number of events using the relation 
\begin{equation}
v_3^\gamma\{\text{PP}\}=  \langle \cos (3(\phi - \psi_{3}^{\text{PP}})) \rangle_{\text{events}} \, ,
\end{equation}
here the participant plane angle $\psi_{3}^{\text{PP}}$ is defined as
\begin{equation}
  \psi_{3}^{\text{PP}} = \frac{1}{3} \arctan 
  \frac{\int \mathrm{d}x \mathrm{d}y \; r^3 \sin \left( 3\phi \right) \varepsilon\left( x,y,\tau _{0}\right) } 
  { \int \mathrm{d}x \mathrm{d}y \; r^3 \cos \left( 3\phi \right) \varepsilon\left( x,y,\tau _{0}\right)}  + \pi/3 \, .
\end{equation}
$\varepsilon$ is the energy density, $r^{2}=x^{2}+y^{2}$, and $\phi$ is
the azimuthal angle in the above equation. The initial triangular eccentricity of the matter density is calculated using 
\begin{equation}
  \epsilon_{3} = -\frac{\int \mathrm{d} x \mathrm{d} y \; r^{3} \cos \left[
  3\left( \phi -\psi_{3}^{\text{PP}}\right) \right] \varepsilon \left(
  x,y,\tau_{0}\right) } {\int \mathrm{d} x \mathrm{d} y \; r^{3} \varepsilon
  \left( x,y,\tau _{0}\right) } \, . 
\end{equation}
$v_3$(PP) and the reaction plane (RP) $v_3$ are shown in figure~\ref{v3} along with the elliptic flow parameter $v_2$ of thermal photons for comparison. We see a  large value of $v_3$(PP) for $p_T <$ 6 GeV/$c$ compared with the elliptic flow results. In addition,  $v_3$(PP) shows similar qualitative nature to the photon $v_2$ where it is small at low $p_T$, then rises with $p_T$, peaks around  2--3 GeV/$c$, and then drops for larger values of $p_T$.

It is well known that the average collision geometry plays a significant role in determining $v_2$ and thus one finds large elliptic flow even with smooth initial conditions and with respect to the reaction plane angle. However, $v_3$ (RP) from individual events is found to be both positive and negative as $\epsilon_3$ is uncorrelated with the RP, and thus after averaging over a large number of events  we find $v_3$(RP)  close to zero.

In case of hadrons, the magnitude of the initial triangular eccentricity $\epsilon_3$ is the main determining factor for the size of the final state $v_3$. However this is not true for photons. From a study of individual events  we see that events with larger early time values of  transverse flow velocity ($v_T$) correlate best with $v_3$ for thermal photons and that conversely the magnitude of $v_3$ does not strongly correlate with the value of $\epsilon_3$. We conclude that fast  $v_T$ builtup driven by fluctuations dominates over the initial geometry effects in determining the $v_3$ of thermal photons (see \cite{csr} for details).

The photon $v_3$ results are expected to be sensitive to the initial parameters of the model calculation and  we study this dependence by changing $\sigma$ and $\tau_0$ from their default values. A larger $\sigma$ results in a smaller triangular flow whereas a larger $\tau_0$ gives larger $v_3$ especially in the region $p_T >$ 2 GeV/$c$.   The initial density distribution becomes smoother for larger $\sigma$ and we get smaller $v_3$. On the other hand for larger $\tau_0$, a large fraction of the high $p_T$ photons with smaller flow velocities are not included in the calculation and we get a much larger $v_3$ at higher $p_T$ (at the expense of significantly reducing the total yield).

\section{Conclusions}
We see a positive and substantial value of the triangular flow parameter $v_3$ of thermal photons at LHC from e-by-e ideal hydrodynamic model. $v_3$(PP) is found to be about half of the value of $v_2$(PP) at $p_T$= 1 GeV/$c$. However, for $p_T> 3$ GeV/$c$ the difference between these two anisotropy parameters is less than 25\%. We know that for hadronic $v_3$ the initial triangular geometry of the overlapping zone dominates over the local fluctuations, however this is not true for photon $v_3$.
We conclude from a study of individual events that unlike in the case of hadron $v_3$ where the magnitue of $\epsilon_3$ is a good predictor for the observed value, photon $v_3$ is driven by a different dynamics having to do with fast buildup of transverse flow due to fluctuations.

\section*{Acknowledgements}
RC gratefully acknowledge he financial support by the Dr. K. S. Krishnan Research Associateship from Variable Energy Cyclotron Centre, Department of Atomic Energy, Government of India. TR is supported by the Academy
researcher program of the Academy of Finland, Project
No. 130472.  We acknowledge the computer facility of the CSC computer centre, Espoo.

\end{document}